\documentclass[twocolumn,trackchanges]{aastex63}
\usepackage{amsmath}
\usepackage{graphicx}
\usepackage{amsfonts}
\usepackage{natbib}
\usepackage{color}
\usepackage{epstopdf}
\usepackage{graphicx}
\usepackage{multirow}
\usepackage{longtable}
\usepackage{rotating}

\shorttitle{Radiation study of CTB 87}
\shortauthors{Gong et al.}

\begin{document}

\title{Multiband Nonthermal Radiative Properties of the Pulsar Wind Nebula CTB 87}

\author{Yunlu Gong}
\affil{Department of Astronomy, School of Physics and Astronomy, Key Laboratory of Astroparticle Physics of Yunnan Province, Yunnan University, Kunming 650091, People's Republic of China; lizhang@ynu.edu.cn, fangjun@ynu.edu.cn}

\author{Liancheng Zhou}
\affil{Department of Astronomy, School of Physics and Astronomy, Key Laboratory of Astroparticle Physics of Yunnan Province, Yunnan University, Kunming 650091, People's Republic of China; lizhang@ynu.edu.cn, fangjun@ynu.edu.cn}

\author{Qi Xia}
\affil{Department of Astronomy, School of Physics and Astronomy, Key Laboratory of Astroparticle Physics of Yunnan Province, Yunnan University, Kunming 650091, People's Republic of China; lizhang@ynu.edu.cn, fangjun@ynu.edu.cn}

\author{Haiyun Zhang}
\affil{Department of Astronomy, School of Physics and Astronomy, Key Laboratory of Astroparticle Physics of Yunnan Province, Yunnan University, Kunming 650091, People's Republic of China; lizhang@ynu.edu.cn, fangjun@ynu.edu.cn}

\author{Jun Fang}
\affil{Department of Astronomy, School of Physics and Astronomy, Key Laboratory of Astroparticle Physics of Yunnan Province, Yunnan University, Kunming 650091, People's Republic of China; lizhang@ynu.edu.cn, fangjun@ynu.edu.cn}

\author{Li Zhang}
\affil{Department of Astronomy, School of Physics and Astronomy, Key Laboratory of Astroparticle Physics of Yunnan Province, Yunnan University, Kunming 650091, People's Republic of China; lizhang@ynu.edu.cn, fangjun@ynu.edu.cn}

\begin{abstract}
The pulsar wind nebula CTB 87 (G74.9+1.2) is one of the sources emitting $\gamma$-rays with energies higher than 10 TeV, as measured by the Very Energetic Radiation Imaging Telescope Array System telescope (VERITAS). In this study, we undertake a reanalysis of the GeV emission from the CTB 87 region, utilising $\sim$16 years of high-energy $\gamma$-ray data collected with the Fermi Large Area Telescope. In the energy range of 0.03--1 TeV, the spectrum can be adequately described by a power-law model with an index of 1.34 $\pm$ 0.18, and the integral energy flux is calculated to be (7.25 $\pm$ 1.36) $\times$ 10$^{-13}$ erg cm$^{-2}$ s$^{-1}$. Based on the multiband data, we have employed a time-dependent model to investigate the non-thermal emission properties of CTB 87. In the model, it is assumed that particles with broken power-law energy distributions are continuously injected into the nebula. This results in multiband non-thermal emission being produced by relativistic leptons via synchrotron radiation and inverse Compton processes. Furthermore, the model suggests an energy of approximately 2.4 PeV for the most energetic particle in the nebula.
\end{abstract}

\keywords{Non-thermal radiation sources - Pulsar wind nebulae - Astronomy data analysis}

\section{Introduction}
\label{sec:intro}
To date, more than 200 very-high-energy (VHE; $>$100 GeV) and ultra-high-energy (UHE; $>$100 TeV) $\gamma$-ray sources have been detected by telescopes with high sensitivity, such as the High Energy Stereoscopic System \citep[H.E.S.S.;][]{2004APh....22..109A}, the High Altitude Water Cherenkov telescope \citep[HAWC;][]{2013APh....50...26A}, and the Large High Altitude Air Shower Observatory \citep[LHAASO;][]{2024ApJS..271...25C}. Pulsar wind nebulae (PWNe) are the most powerful particle accelerators within our Galaxy, as evidenced by the detection of VHE photons emitted from them \citep{2024ApJ...967..127Z,2024ApJS..271...25C}. The LHAASO collaboration has detected $\gamma$-ray emission with energies up to 1.1 peta-electron volts in the Crab Nebula \citep{2021Sci...373..425L}. In the subsequent phases of PWNe evolution, the relativistic particle population will be injected into the surrounding interstellar medium and may contribute to the cosmic ray electron-positron population \citep{2009PhRvD..80f3005M,2023ApJ...945....4E}. Consequently, multiwavelength studies of $\gamma$-ray sources are imperative to explore the origin of the highest energy galactic cosmic rays and to comprehend their particle acceleration, radiation, and cooling \citep{2023MNRAS.526..193G}.

The center-filling morphology, the undetected shell, the linearly polarized radio flux, and the early X-ray observations from Einstein's satellites all point to the classification of CTB 87 as a plerionic type supernova remnant \citep{1975AJ.....80..437D,1975A&A....38..461D,1980ApJ...241L..19W}. It appears that there is a spectral break (spectral index changes from -0.29 to -1.08) in the radio spectrum of CTB 87 above 10 GHz. This phenomenon may be attributed to the absence of large-scale emission and the limited sensitivity of high-frequency radio continuum observations \citep{1987A&AS...69..533M,2006A&A...457.1081K}. However, observations of the 2 cm wavelength radio data indicate that any notable high-frequency spectral bending or breaking should occur well above 18 GHz \citep{2022A&A...668A..39R}. In an evolutionary simulation of the PWN, \cite{2020MNRAS.496..723K} suggest that the same supernova explosion about 18,000 years ago produced a complex system, CTB 87, with an ejecta mass of $\sim12\mathrm{~M}_\odot$ and an explosion energy of $\sim7 \times 10^{50}$ erg. Recently, \cite{2024MNRAS.528.6761L} used the 500-meter Aperture Spherical radio Telescope to make the first discovery of the radio pulsar PSR J2016 + 3711 in CTB 87, with a pulse significance of $\sim10.8\sigma$, thus confirming the compact nature of the X-ray point source.

Observations in X-rays with the Chandra X-Ray Observatory revealed a putative pulsar, CXOU J201609.2+371110, and an offset of $\sim$100 arcsec was found between the X-ray and radio emission peaks \citep{2013ApJ...774...33M}. The CTB 87 is an evolved ($\sim$5-28 kyr) PWN, on the reasonable assumption that the pulsar born at the radio peak is now at the X-ray peak \citep{2013ApJ...774...33M}. In a deep XMM-Newton observation, \cite{2020MNRAS.491.3013G} found no evidence for thermal X-ray emission from a surrounding supernova remnant and concluded that the morphology and spectral properties were consistent with a $\sim$20 kyr old PWN expanding into a wind-blown bubble. Furthermore, \cite{2020MNRAS.491.3013G} estimated a period of 0.065 s, a period derivative of $0.51\times10^{-13} \mathrm{s} \ \mathrm{s}^{-1}$, and a surface magnetic field of $3.8\times10^{12}$ G, based on the pulsar's characteristic age (11.5 kyr) and the predicted spin-down energy loss ($\dot{E}=7.5\times10^{36} \mathrm{erg\ s}^{-1}$).

The $\gamma$-ray emission from CTB 87 was first detected by the MILAGRO Gamma-Ray Observatory in the direction of MGRO J2019+37, and was subsequently resolved by VERITAS into two emission components. Of these, VER J2016+371 is spatially consistent with the peak of the radio emission from CTB 87 \citep{2007ApJ...658L..33A,2014ApJ...788...78A}. In the Fermi Large Area Telescope (Fermi-LAT) analysis, \cite{2018ApJ...861..134A} used two power-law source models to fit the data and found that the spectra of the two sources were significantly different, with the source at the radio position of CTB 87 being weaker and harder. Combining the position of VER J2016+371 with the spectrum of the Fermi-LAT emission, \cite{2018ApJ...861..134A} concluded that the Fermi emission at CTB 87 and the TeV emission at VER J2016+371 originated from the same place, but that the contribution of QSO J2015+371 (TS $\simeq$ 1087) to the VHE emission could not be excluded. \cite{2021ApJ...911..143A} used HAWC data to analyse the MGRO J2019+37 region, distinguishing two sources: HAWC J2019+368 and HAWC J2016+371. However, there was no significant detection of HAWC J2016+371 in any of the individual energy bins.

In a PWN scenario, \cite{2016MNRAS.460.3563S} showed that the Maxwellian distribution of electrons and the broken power-law (BPL) distribution of electrons in low magnetic fields can explain the multiband data at VER J2016+371, while the BPL distribution of electrons is incompatible with the data at MeV-GeV energies. Based on HI absorption measurements and CO observations of potentially relevant molecular material, the distance of CTB 87 was estimated to be 6.1 $\pm$ 0.9 kpc by \cite{2003ApJ...588..852K} and later confirmed the \cite{2018ApJ...859..173L} study of a nearby molecular cloud complex.

In this paper, we analyse the GeV-TeV emission observed by Fermi-LAT for CTB 87 in greater detail, and investigate the multiband non-thermal emission properties using a time-dependent model based on the latest multiband data. In Section~\ref{sec:data}, we present the detailed Fermi-LAT data analysis process and results. In Section~\ref{analysis}, we describe the model in detail. In Section~\ref{discussion}, the results of the model calculations and discussion are given. Finally, a summary is presented in Section~\ref{summary}.

\begin{figure*}[!hbt]
	\includegraphics[width=\textwidth]{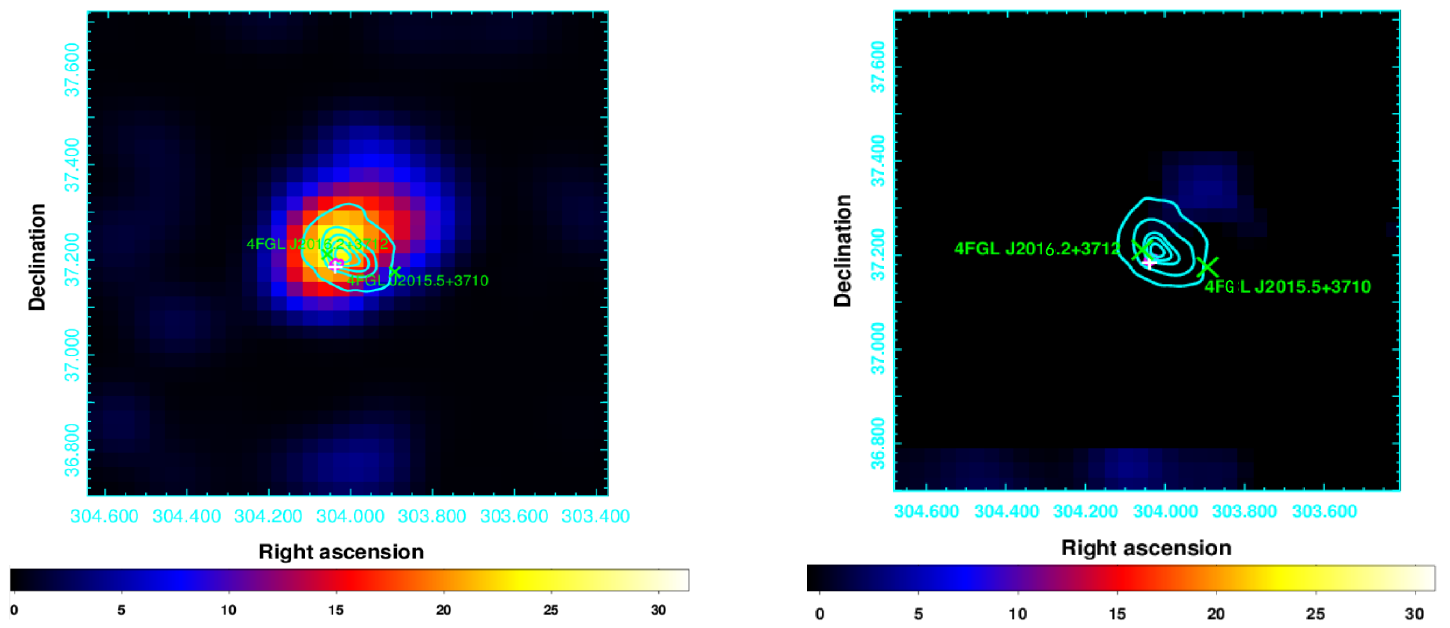}
    \caption{TS maps of 30 GeV-1 TeV with $0_{\cdot}03^{\circ}$ pixel size centered at the best-fit position of CTB 87, and the TS maps of $1_{\cdot}0^{\circ} \times 1_{\cdot}0^{\circ}$ are smoothed with a Gaussian kernel of $\sigma = 0_{\cdot}2^{\circ}$. The cyan contours indicate the 14.7 GHz emission detected by the 100-metre Effelsberg telescope \citep{2022A&A...668A..39R}. The purple contours represent the X-ray image of CTB 87 from a 125 ks observation with XMM-Newton in the 0.3-10 keV energy range \citep{2020MNRAS.491.3013G}. The white plus shows the position of the radio pulsar PSR J2016 + 3711 \citep{2024MNRAS.528.6761L}. The 4FGL-DR4 sources are indicated by the green crosses. Left panel: TS map for the data above 30 GeV with the diffuse backgrounds and 4FGL-DR4 sources subtracted, except for 4FGL J2016.2+3712. Right panel: TS map after subtracting all sources containing 4FGL J2016.2+3712.
    }
    \label{fig1}
\end{figure*}

\section{Fermi-LAT Data Analysis}
\label{sec:data}

\subsection{Data Reduction}
\label{data1}
The Fermi-LAT is uniquely sensitive to high-energy $\gamma$-ray observations, scanning the entire sky every 3 hours \citep{2009ApJ...697.1071A}. The latest Pass 8 data taken from 2008 August 4 (mission elapsed time 239557417) to 2024 July 28 (mission elapsed time 743821401) are collected to study the GeV-TeV emission around the CTB 87 region. Recently, \cite{2024MNRAS.528.6761L} attempted to search for $\gamma$-ray pulses in the LAT data using ephemerides derived from radio observations, but unfortunately no pulse signals were detected. Therefore, in order to avoid as much as possible the effects of pulsar emission, an energy range between 30 GeV and 1 TeV was chosen \citep{2021Natur.594...33C}. This will allow better detection of the $\gamma$-ray emission from the wind nebula itself. Following the recommendations of the LAT team, we used data with an event class of `source' ({\tt evclass = 128 \& evtype = 3}) and the instrumental response function {\tt ``P8R3\_SOURCE\_V3''}. In addition, we have excluded events with a zenith angle greater than $90^{\circ}$ to reduce the contamination from the Earth's limb, and events with a ``bad" quality flag. In the binned maximum likelihood analysis, we considered data within a $20^{\circ} \times 20^{\circ}$ region of interest (ROI) centred on the position of CTB 87. Then, the Fermi-LAT Fourth Source Catalog \citep[4FGL-DR4;][]{2023arXiv230712546B} and the script {\tt make4FGLxml.py} was used to generate the source model files, which include diffuse Galactic interstellar emission ({\tt gll\_iem\_v07.fits}) and isotropic emission ({\tt iso\_P8R3\_SOURCE\_V3\_v1.txt}) templates. The python scripts are standard Fermi-LAT analysis tools provided by NASA\footnote{https://fermi.gsfc.nasa.gov/ssc/data/analysis/user/}. The normalizations and spectral parameters of the sources from the center of ROI of 5$^{\circ}$, together with the normalizations of the isotropic and Galactic components, are left free.

\subsection{Spatial Analysis and Spectral Analysis}
The detection significance of a new source having one additional free parameter, for example, can be calculated as the square root of the test statistic (TS), defined as $\mathrm{TS}=-2\log{(\mathcal{L}_{0}/\mathcal{L})}$, where $\mathcal{L}_{0}$ is the likelihood of the null hypothesis (background only) and $\mathcal{L}$ is the maximum likelihood of the alternative hypothesis (background + new source). In order to obtain a clear understanding of the potential emissions in the vicinity of the CTB 87 area, a TS map was generated by executing the {\tt gttsmap} command in the 0.03--1 TeV range. The left panel of Fig.~\ref{fig1} shows a significant excess of $\gamma$-ray emission in the direction of CTB 87, where all catalogue sources except 4FGL J2016.2+3712 have been removed. For radio pulsars, the contours of the radio and X-ray emission are in good spatial agreement with the GeV-TeV emission. This finding is further corroborated by related studies by \cite{2014ApJ...788...78A} and \cite{2018ApJ...861..134A}, which demonstrate that the emission from VER J2016+371 is spatially consistent with both the peak of the radio emission and the GeV emission. After subtracting the emission from source 4FGL J2016.2+3712, no significant residual $\gamma$-ray emission is detected at the position of CTB 87 (see right panel of Fig.~\ref{fig1}), thus suggesting that the $\gamma$-ray emission most likely originated from this region. Next, the best-fit position of CTB 87 calculated by the {\tt gtfindsrc} command is R.A., decl. = $304.05^{\circ}$, $37.21^{\circ}$ with 1$\sigma$ error radius of $0.02^{\circ}$, which is marked as SrcX for all subsequent analyses. The angular distances between the best-fitting position and 4FGL J2016.2+3712 is $0.01^{\circ}$.

Similar to the analysis of \cite{2018ApJ...861..134A}, we used two power-law point sources (SrcX and 4FGL J2015.5+3710) to model the emission due to the proximity of the blazar 4FGL J2015.5+3710 to the centre of the ROI. The calculations show that the blazar 4FGL J2015.5+3710 has a TS value of 2.02 in the energy range 0.03--1 TeV, which allows us to neglect its emission contribution in the SrcX region compared to the results (TS $\sim$ 1087) of the analysis by \cite{2018ApJ...861..134A}. The TS value for SrcX was computed to be 37.21, corresponding to a significance level of 5.1$\sigma$ for four degrees of freedom. The integral photon flux is calculated to be (1.62 $\pm$ 0.55) $\times$ 10$^{-11}$ photons cm$^{-2}$ s$^{-1}$ with spectral index of 1.34 $\pm$ 0.18. In addition, the spatial extension of SrcX was investigated by utilising uniform disc and 2D Gaussian templates. The radius and $\sigma$ of these templates were varied within the range of $0.01^{\circ}$ to $0.5^{\circ}$, with a step size of $0.01^{\circ}$. The significance of the source extension is quantified as the test statistic $\mathrm{TS_{ext}}=2{(\ln\mathcal{L}_{ext} - \ln\mathcal{L}_{pt})}$, which compares the overall maximum likelihood of the extended template ($\mathcal{L}_{ext}$; alternative hypothesis) with that of the point-like source model ($\mathcal{L}_{pt}$; null hypothesis). A study of the parameter distributions suggests that the value $\mathrm{TS_{ext}}$ = 16 ($\sqrt{\mathrm{TS_{ext}}} \simeq 4\sigma$) is an appropriately low threshold for selecting the extended template as the preferred model for the source \citep{2012ApJ...756....5L}. The results of the tests with the highest TS values in the different models are shown in Table~\ref{tab1}, and no significant improvement was found by calculating TS$_\mathrm{{ext}}$ values ($<$1). For this reason, the point source template of the SrcX will be chosen for the following analysis. \cite{2018ApJ...861..134A} also selected a point source template with a power law form to fit the TeV emission data for VER J2016+371.

The spectral energy distribution (SED) was generated in the 0.03--1 TeV energy band for SrcX. The data was divided into four equal logarithmic energy bins, and the binned likelihood method was repeated for each energy bin, with the spectral normalisations of sources within $5.0^{\circ}$ around SrcX and the two diffuse backgrounds left free. In instances where the TS value is less than 4.0 (e.g. the third bin), an upper limit is calculated with a significance level of 95\%. The resulting SED of SrcX is displayed in Fig.~\ref{fig5}, and the corresponding data are listed in Table~\ref{tab2}.

\begin{table*}
   \centering
   \caption{Spatial Properties for SrcX between 30 GeV and 1 TeV}
   \label{tab1}
   \renewcommand\arraystretch{1.0}
  \setlength{\tabcolsep}{3.3mm}
   \begin{tabular}{ccccccc} 
      \hline \hline
Spatial Template &Sources& Radius ($\sigma$) & Spectral Index & Photon Flux & TS Value & Degrees   \\
 &	 &  &  & 10$^{-11}$ ph cm$^{-2}$ s$^{-1}$ & & of Freedom  \\\hline
Two points & SrcX & ...  & 1.34 $\pm$ 0.18 & 1.62 $\pm$ 0.55 & 37.21 & 8 \\
 &	J2015.5+3710 & ...  & 2.61 $\pm$ 0.26 & 0.35 $\pm$ 0.31 & 2.02 &  \\
Disk+Point &SrcX	&  $0.05^{\circ}$ & 1.29 $\pm$ 0.30 & 1.46 $\pm$ 0.51 & 37.67 & 9 \\
     &J2015.5+3710	&  ... & 2.50 $\pm$ 0.25 & 0.61 $\pm$ 0.37 & 6.74 &  \\
Gaussian+Point &SrcX	&  $0.06^{\circ}$ & 1.33 $\pm$ 0.31 & 1.69 $\pm$ 0.62& 36.49 & 9 \\
  &J2015.5+3710	&  ... & 2.50 $\pm$ 0.25 & 0.59 $\pm$ 0.39 & 5.12 &  \\\hline
\multicolumn{4}{l}{}
   \end{tabular}
\end{table*}

\begin{table}
   \centering
   \caption{Fermi-LAT Spectral Points for SrcX Fit with a Power Law}
   \label{tab2}
   \renewcommand\arraystretch{1.0}
  \setlength{\tabcolsep}{2.0mm}
   \begin{tabular}{cccc} 
      \hline \hline
E & Band & E$^2$dN(E)/dE & TS Value  \\
 GeV&	GeV &  [10$^{-12}$ erg cm$^2$ s$^{-1}$]& \\\hline
 46.50&	30.00--72.08 & 0.51$\pm$ 0.29 & 10.02\\
 111.74&	72.08--173.20 & 1.41 $\pm$ 0.96 & 19.19\\
 263.04& 173.20--416.17 & $<$0.66 & 0.00\\
 645.12&416.17--1000.0 & 2.73 $\pm$ 2.34 & 8.01\\\hline
   \end{tabular}
\end{table}

\section{Model Description}
\label{analysis}
Here, in order to understand the multiband non-thermal emission properties of PWN CTB 87, a time-dependent one-zone model is employed \citep{2024MNRAS.529.3593Z,2024ApJ...972...84X}. In this model, the PWN is conceptualised as a sphere in which energetic leptons (electrons and positrons) are continuously injected. The evolution of the lepton distribution $N(\gamma,t)$ can be obtained from the diffusion equation \citep[e.g.][]{2010A&A...515A..20F,2012MNRAS.427..415M}
\begin{equation}
\frac{\partial N(\gamma,t)}{\partial t}=\frac{\partial}{\partial\gamma}[\dot{\gamma}(\gamma,t)N(\gamma,t)]-\frac{N(\gamma,t)}{\tau(\gamma,t)}+Q(\gamma,t) ,
\label{eq1}
\end{equation}
where $\dot{\gamma}(\gamma,t)$ is the summation of the energy losses due to synchrotron, (Klein-Nishina) inverse Compton, self-synchrotron Compton, and adiabatic expansion. $\tau(\gamma,t)$ represents the escape time of particles via Bohm diffusion \citep{2008ApJ...676.1210Z}, and $Q(\gamma,t)$ is the injection of particles per unit energy (or Lorentz factor) and unit volume in a certain time.

If the pulsar ephemeris is known, then the spin-down luminosity can be determined by
\begin{equation}
L(t)=4\pi^2I \frac{P}{\dot{P}^3},
\label{eq2}
\end{equation}
where $P$ and $\dot{P}$ are the period and period derivative, respectively, in the pulsar ephemeris. Different pulsars have different moment of inertia. $I = 2MR^2/5$ is the moment of inertia of the pulsar, and the canonical values of $R = 10 km, M = 1.4 M_{\bigodot}$ lead to $I = 10^{45}\text{g cm}^2$ \citep{2006ARA&A..44...17G,2014JHEAp...1...31T,2020PhDT.........2G}. In this case, the moment of inertia of the pulsar is assumed to be $10^{45}\text{g cm}^2$. Moreover, an additional parameter derived from the pulsar ephemeris is the characteristic age, which is written as
\begin{equation}
\tau_\mathrm{c}\equiv\frac{P}{2\dot{P}}=(\tau_0+t) \frac{n-1}{2},
\label{eq3}
\end{equation}
where $n$ is the breaking index and a typical value is considered to be 3 (corresponding to a dipole spin-down rotator). $t$ is the true age of the pulsar. $\tau_0$ is the initial spin-down time-scale of the pulsar,
\begin{equation}
\tau_0=\frac{P_0}{(n-1)\dot{P}_0}=\frac{2\tau_\mathrm{c}}{n-1}-t_\mathrm{age},
\label{eq4}
\end{equation}
where $P_0$ and $\dot{P}_0$ are the initial period and its first derivative, respectively. The normalization constant $\mathcal{Q}_{0}(t)$ is determined using the injection luminosity $L(t)$, which is given by \cite{2006ARA&A..44...17G}
\begin{equation}
L(t)=L_0\left(1+\frac{t}{\tau_0}\right)^{-\frac{n+1}{n-1}},
\label{eq5}
\end{equation}
where $L_0$ is the initial luminosity.

The spin-down luminosity of an energetic pulsar is continuously transferred to particles and magnetic field. At any point in the evolution, only a small fraction of the rotational energy drives the magnetic field, so the magnetic energy fraction $\eta$ is relatively small. The time-evolving magnetic field strength in PWN can be estimated by \citep{2010ApJ...715.1248T,2012MNRAS.427..415M}
\begin{equation}
B(t)=\sqrt{\frac{3(n-1)\eta L_0\tau_0}{R_{\text{PWN}}^3(t)}\left[1-\left(1+\frac{t}{\tau_0}\right)^{-\frac{2}{n-1}}\right]},
\label{eq6}
\end{equation}
Assume that the nebula has not encountered a reverse shock from a supernova remnant and that it is still in the free expansion phase. The radius of the nebula is \citep{2001A&A...380..309V,2003A&A...404..939V,2014JHEAp...1...31T}
\begin{equation}
R_{\mathrm{PWN}}(t)=0.84\left(\frac{L_0t}{E_0}\right)^{1/5}\left(\frac{10E_0}{3M_{\mathrm{ej}}}\right)^{1/2}t,
\label{eq7}
\end{equation}
where $E_0$ and $M_{\mathrm{ej}}$ are the kinetic energy of the supernova ejecta and the ejected mass, respectively. We assume that leptons are continuously injected into the PWN with a broken power law distribution \citep{1984ApJ...283..710K,2020MNRAS.498.4901F}
\begin{equation}
Q(\gamma,t)=Q_0(t)\left\{\begin{array}{l}\left(\frac{\gamma}{\gamma_\mathrm{b}}\right)^{-\alpha_1}\quad\text{if} \ \gamma\leq\gamma_\mathrm{b} ,\\\left(\frac{\gamma}{\gamma_\mathrm{b}}\right)^{-\alpha_2}\text{if} \ \gamma_\mathrm{b}<\gamma\leq\gamma_\mathrm{max} .\end{array}\right.
\label{eq8}
\end{equation}
where $\gamma_\mathrm{b}$ represents the break energy, while $\alpha_1$ and $\alpha_2$ are the spectral indices. The normalization constant $Q_0(t)$ is given by \citep{2014JHEAp...1...31T}
\begin{equation}
(1-\eta)L(t)=\int_{\gamma_{\min}}^{\gamma_{\max}}\gamma m_{e}c^{2}Q(\gamma,t)\mathrm{d}\gamma,
\label{eq9}
\end{equation}
The minimum energy at injection is a free parameter, we assume it to be $\gamma_{\min}=1$ \citep{2012MNRAS.427..415M,2016MNRAS.459.3868M}. Additionally, the maximum energy at injection is determined by \citep{2014JHEAp...1...31T}
\begin{equation}
\gamma_{\max}(t)=\frac{\varepsilon e\kappa}{m_{\mathrm{e}}c^2}\left(\frac{\eta L(t)}{c}\right)^{1/2}.
\label{eq10}
\end{equation}
where $e$ and $m_{\mathrm{e}}$ are the electron mass and charge, respectively. The magnetic compression ratio $\kappa$ depends on the strength of the relativistic shock (and hence the magnetization parameter $\sigma$). In order to satisfy the flow and pressure boundary conditions at the outer edge of the PWN, models for the structure of the Crab-like require $\sigma \ll 1$ (strong shocks; $\kappa \sim 3(1-4\sigma) \sim3$) just behind the termination shock \citep{1984ApJ...283..694K,1984ApJ...283..710K,2003ApJ...593.1013S,2005MNRAS.358..705B,2009ASSL..357..451D}. Here, we assume $\kappa$ to be 3 \citep{2012MNRAS.427..415M,2022JHEAp..36..128M}. To ensure that the particles are confined, the Larmor radius $R_{\mathrm{L}}$ needs to be smaller than a fraction of the radius of the termination shock $R_{\mathrm{s}}$ ($R_{\mathrm{L}}=\varepsilon R_{\mathrm{s}}$). For this reason, the fractional size ($\varepsilon$) of the radius of the shock should be less than 1. Typically, to calculate the maximum energy of the particle for PWNe, $\varepsilon$ is used in the range of 0.2 to 1/3 \citep{2014JHEAp...1...31T,2020MNRAS.498.4901F}.

\begin{figure}[!hbt]
	\includegraphics[width=.49\textwidth]{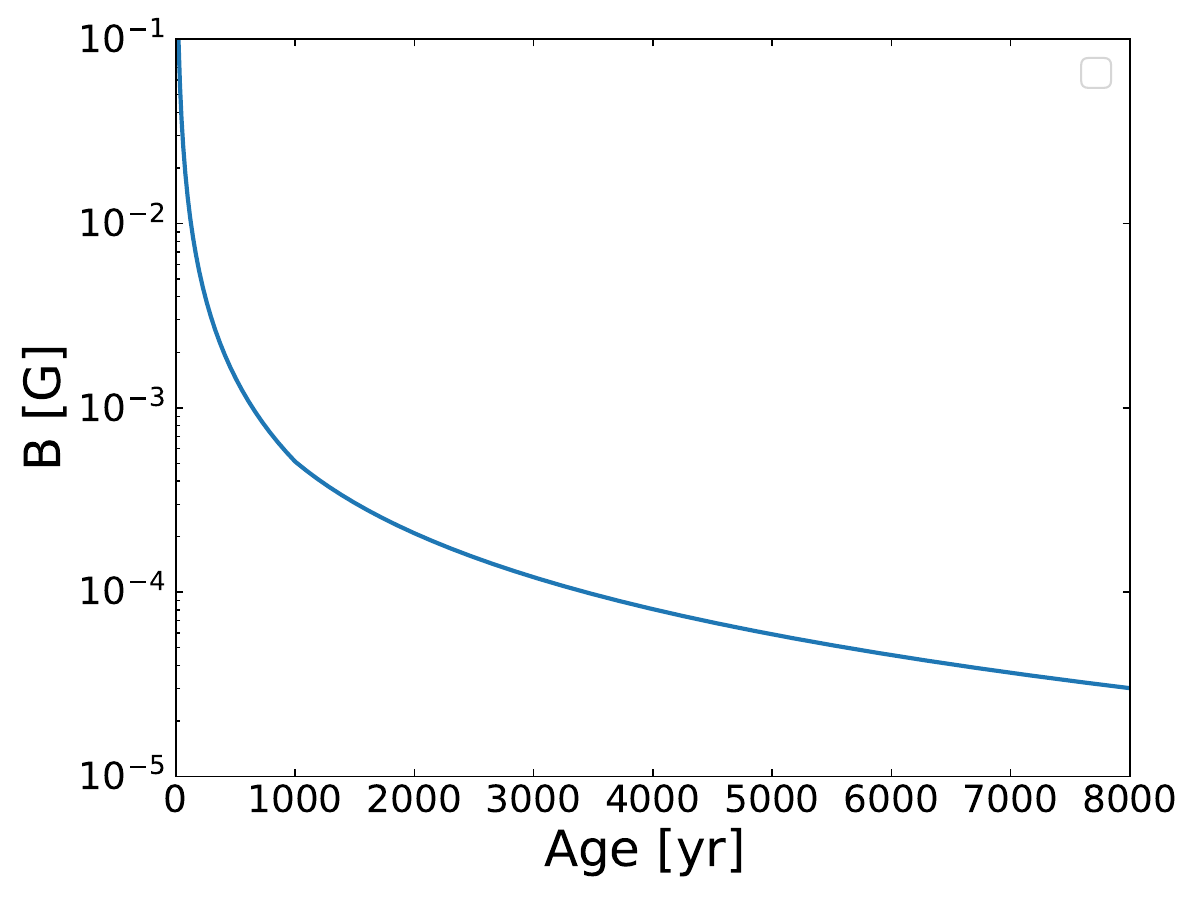}
    \caption{The result of the variation in magnetic field strength over time with $\eta = 0.09$.
    }
    \label{fig2}
\end{figure}

\begin{figure}[!hbt]
	\includegraphics[width=.49\textwidth]{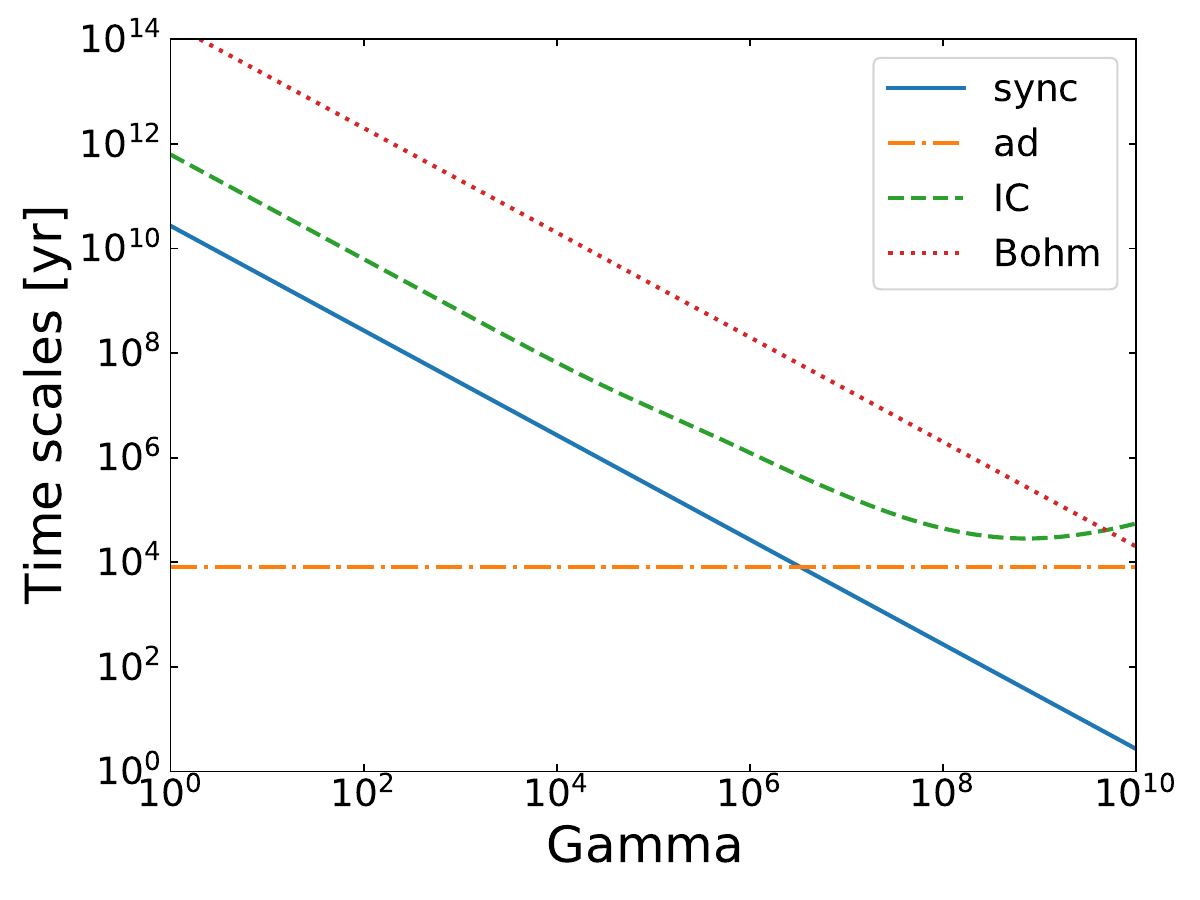}
    \caption{Cooling times for synchrotron radiation, adiabatic loss, inverse Compton scattering, and the escape time for Bohm diffusion at $t_{\mathrm{age}}=8 \ \mathrm{kyr}$.
    }
    \label{fig3}
\end{figure}

\begin{figure}[!hbt]
	\includegraphics[width=.49\textwidth]{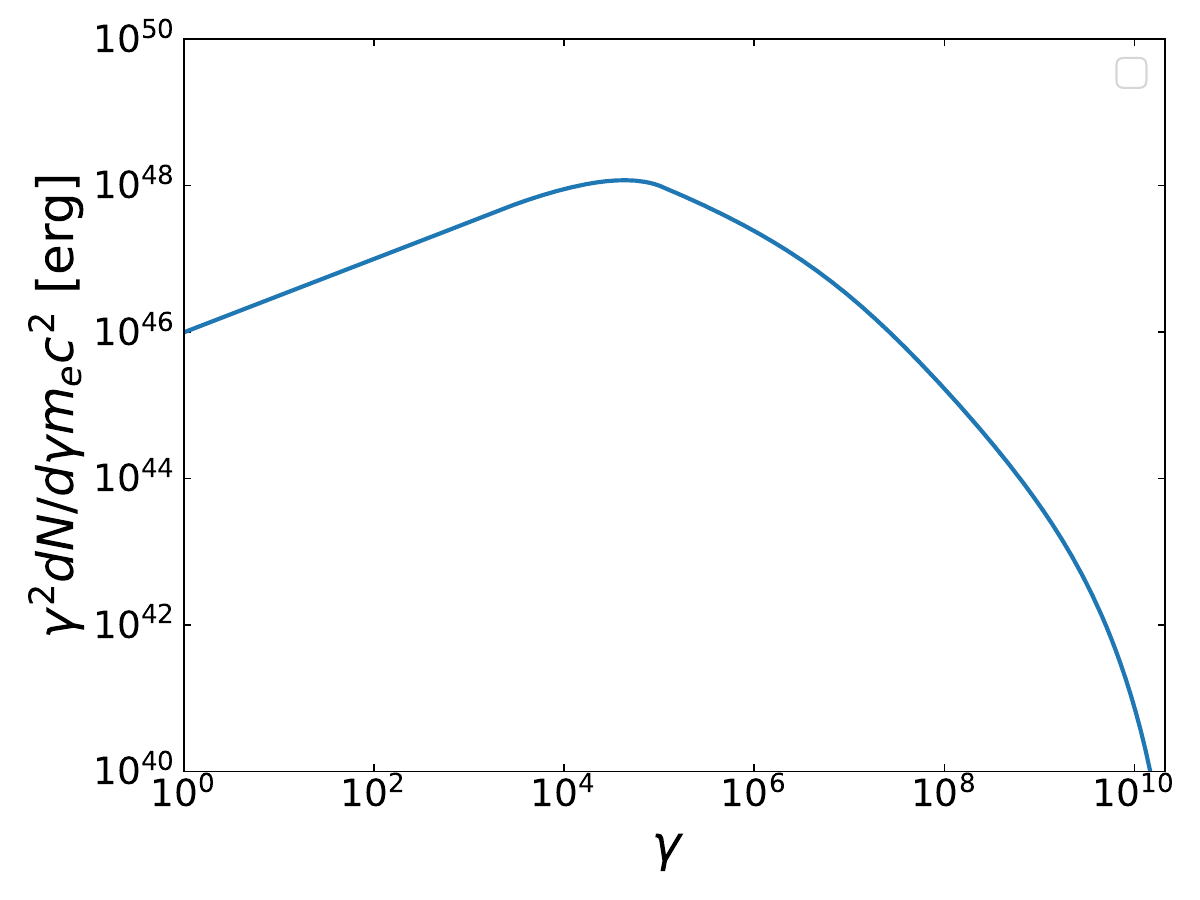}
    \caption{Particle distributions of the leptons in the nebula on $\gamma$ for the BPL models at $t_{\mathrm{age}}=8 \ \mathrm{kyr}$.
    }
    \label{fig4}
\end{figure}

\begin{figure}[!hbt]
	\includegraphics[width=.49\textwidth]{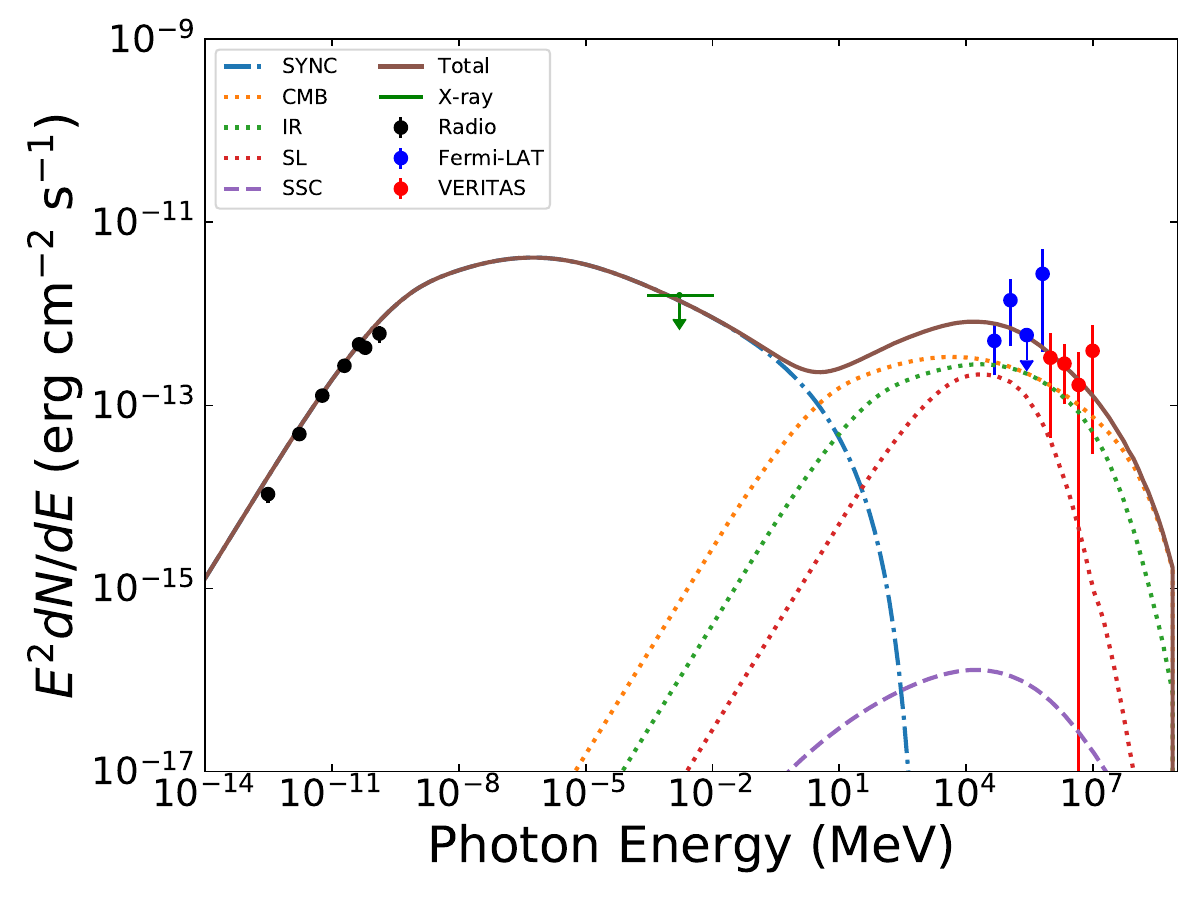}
    \caption{SED for the CTB 87 obtained by fitting our model with $t_{\mathrm{age}}=8 \ \mathrm{kyr}$. The data points are obtained from \cite{2006A&A...457.1081K}, \cite{2012RaSc...47.0K04L}, and \cite{2020MNRAS.496..723K} for radio band; \cite{2020MNRAS.491.3013G} for X-ray band; Analysis results of Section~\ref{sec:data} for Fermi band; \cite{2018ApJ...861..134A} for TeV band.
    }
    \label{fig5}
\end{figure}

\section{Results and Discussion}
\label{discussion}
In order to model the multi-wavelength properties of CTB 87, we have drawn upon a range of data sources. Specifically, the radio flux densities between 74 and 32 000 MHz were taken from the Effelsberg 100-m Radio Telescope \citep{2020MNRAS.496..723K}, while the 74 MHz and 408 MHz flux densities were taken from the Very Large Array Low-frequency Sky Survey Redux \citep{2012RaSc...47.0K04L} and the CGPS SNR catalogue \citep{2006A&A...457.1081K}, respectively. With regard to the X-ray data, the total flux 16.01 $\times$ 10$^{-13}$ erg cm$^{-2}$ s$^{-1}$ obtained by an absorbing power-law model was considered to be an upper limit, and the size of the nebular emission observed by XMM-Newton was taken to be 171 arcsec $\times$ 129 arcsec \citep{2020MNRAS.491.3013G}. For the GeV-TeV data, the results of the analysis are employed, with the effects of the blazar 4FGL J2015.5+3710 emission being neglected in section~\ref{sec:data}. The fluxes observed at TeV energies taken from VERITAS and fit with a power law gives the parameters $N_0=(2.8\pm 1.2_{\mathrm{stat}}) \times 10^{-17} \mathrm{GeV}^{-1} \mathrm{cm}^{-2} \ \mathrm{s}^{-1} \ \mathrm{at} \ E_{0} = 2510 \ \mathrm{GeV}$ and a spectral index of $2.1\pm0.8_{\mathrm{stat}}$ \citep{2018ApJ...861..134A}. The distance to the source was considered to be 6.1 kpc according to \cite{2003ApJ...588..852K}. The central pulsar in CTB 87 (PSR J2016+3711) is observed in radio to have a period of 50.81 ms, and a first period derivative of $7.2\times10^{-14}\text{s s}^{-1}$, implying a characteristic age of $\tau_c\sim11.1\mathrm{~kyr}$ \citep{2024MNRAS.528.6761L}.

In a PWN scenario, multiband non-thermal emission is produced through synchrotron radiation and inverse Compton scattering (ICS) of relativistic leptons. For the ICS process, three interstellar photon fields are considered: the cosmic microwave background (CMB) photons, the galactic infrared (IR) photons, and the star light (SL) photons. The energy densities U$_\mathrm{{CMB}}$ = 0.25 eV cm$^{-3}$, U$_\mathrm{{IR}}$ = 0.23 eV cm$^{-3}$, and U$_\mathrm{{SL}}$ = 0.5 eV cm$^{-3}$, and temperatures T$_\mathrm{{CMB}}$ = 2.7 K, T$_\mathrm{{IR}}$ = 32.0 K, and T$_\mathrm{{SL}}$ = 3000.0 K for the CMB, IR, and SL photons are used in the calculation. The energy density of the field is obtained using the radiative transfer model for the Milky Way, following the work of \cite{2017MNRAS.470.2539P}. Furthermore, we set the kinetic energy of the supernova ejecta to $E_0=10^{51} \mathrm{erg}$ with an ejected mass of $M_{\mathrm{ej}}=15.0 \ M_{\odot}$. We assume that leptons are continuously injected into the PWN with a broken power law distribution.

There are more than a dozen parameters involved in the use of the model to reproduce the multi-wavelength observations of the PWN. To fit the observed SEDs, we divided the model parameters into two parts: measured/assumed and fitted parameters. The measured/assumed parameters include pulsar and supernova parameters and soft photon parameters (e.g. period $P$, the distance d, the braking index n, pulsar age t$_{age}$, the initial spin-down power $L_0$, $E_0$, and $M_{\mathrm{ej}}$). In addition, the spectral parameters $E_{\mathrm{b}}$, $\alpha_{1}$, and $\alpha_{2}$, the magnetic fraction $\eta$, and shock radius fraction $\varepsilon$ are considered as fitted parameters. To obtain the best fitting values during the fitting process, the Levenberg-Marquardt method of the $\chi^2$ minimization fitting procedure was used \citep{1992nrfa.book.....P,2015MNRAS.451.3145Z,2023ApJ...943...89Z}. The reduced $\chi^2$ value was calculated to be 2.51.

\cite{2013ApJ...774...33M} consider CTB 87 to be an evolved ($\sim$ 5-28 kyr) PWN based on X-ray observations, so we consider the system to be 8 kyr old. With this age, the initial spin-down timescale $\tau_{0}=3.1\times10^{3} \ \mathrm{yr}$ and the initial luminosity $L_0=1.76\times10^{39} \mathrm{erg} \ \mathrm{s}^{-1}$ are estimated. An interesting phenomenon in the study of the multiband nonthermal radiation properties of some PWNe is that the equipartition of magnetic and particle energies is not followed, with the magnetic energy density typically being much lower than the particle energy density \citep{2014JHEAp...1...31T,2018A&A...609A.110Z}. Here the magnetic energy ratio $\eta$ is estimated to be 0.09, which is consistent with other evolved PWNe and suggests that the magnetic energy has been converted to particle energy during its long life cycle. As illustrated in Fig.~\ref{fig2} and equation~\ref{eq6}, the magnetic field strength in the nebula is estimated to be 11.07 $\mu$G at 8000 yr, which is in agreement with the results (7.0 $\mu$G) obtained by \cite{2016MNRAS.460.3563S} using a BPL-type electron distribution. However, \cite{2013ApJ...774...33M} estimated the magnetic field strength to be 55 $\mu$G, which is not a very appropriate value as the magnetic field of evolved PWN is generally thought to be much smaller ($\sim$10 $\mu$G) than that of young PWN like the Crab Nebula ($\sim$110 $\mu$G) \citep{2021Sci...373..425L}. At the same time, the radius was determined to be 19.5 pc, which is marginally larger than that ascertained through ($\sim$15.0 pc) radio observations \citep{2020MNRAS.496..723K,2022A&A...668A..39R}.

The cooling time scales for the adiabatic loss, the synchrotron radiation, the ICS, and the escape time scale due to the Bohm diffusion for the particles with different Lorentz factors are shown in Fig.~\ref{fig3}. The results show that adiabatic loss is the dominant process cooling the particles when $\gamma<3.3 \times 10^{6}$, that synchrotron radiation is the dominant process when $3.3 \times 10^{6} < \gamma <4.6 \times 10^{9}$, and that the particles rapidly escape from the nebula through the Bohm diffusion when $\gamma\sim 10^{10}$. Fig.~\ref{fig4} presents the electron spectrum of the nebula at $t_{\mathrm{age}}=8 \ \mathrm{kyr}$, with a clear break at $\gamma_{\mathrm{b}}=1\times 10^5$. As shown in Fig.~\ref{fig5}, multiband data are well reproduced by the synchrotron, the synchrotron self-Compton (SSC) and the ICS off the CMB, IR background and SL. In our model, the low-energy spectral index has a value of $\alpha_{1}\sim1.51$, the high-energy index is $\alpha_{2}\sim2.43$, the energy break is $E_{\mathrm{b}}\sim5.11\times10^{4} \mathrm{MeV}$, and the fractional size of the radius of the shock is $\varepsilon \sim 1/3$. Furthermore, the maximum Lorentz factor is estimated to be $4.72 \times 10^9$, which means that particles injected into the PWNe can be accelerated up to $\sim$2.41 PeV. As demonstrated in Fig.~\ref{fig5}, we can see that the ICS off the CMB and IR photons are the main contributors to the high energy spectrum, with the contributions from the CMB and IR being almost equal in the 43 GeV-3 TeV energy range, otherwise the CMB is larger than the IR. Additionally, the contribution of the ICS off SL is also clear, while the contribution of the SSC process is not very important.

Observations made by LHAASO have revealed the presence of $\gamma$-ray emission at PeV energies in the Crab Nebula, thereby indicating the potential existence of a PeV accelerator within the nebula. Assuming the particles injected into the Crab Nebula have a spectrum of a BPL with $\alpha_{1}=1.61,\ \alpha_{2}=2.56$, $\gamma_\mathrm{b}=2\times10^{6}$, $\varepsilon = 0.28$, and $\eta = 0.02$, the model can reproduce the detected fluxes from radio to UHE $\gamma$-ray \citep{2021RAA....21..286W}. The two indices and the initial luminosity ($\sim3.1\times10^{39}\mathrm{erg} \ \mathrm{s}^{-1}$) are similar to those used in this paper for CTB 87 powered by PSR J2016+3711. However, PWN CTB 87 has a systematic age about eight times that of the Crab Nebula, and the spin-down energy is injected into the nebula over a longer period of time than in the Crab Nebula \citep{2014JHEAp...1...31T}. Consequently, the $\gamma$-ray production capability of CTB 87 is comparable to that of the Crab Nebula.

It is noteworthy that both VER J2016+371 and 4FGL J2016.2+3712 are spatially close to the blazar 4FGL J2015.5+3710, which suggests a potential relationship between the energetic emission and 4FGL J2015.5+3710. Based on spatial correlations and observed variability in the $\gamma$-ray and radio bands, \cite{2012ApJ...746..159K} suggest that high-energy $\gamma$-ray emission is associated with nearby blazar B2013+370. However, the VHE $\gamma$-ray emission from this blazar has not yet been observed in VERITAS and HAWC, and its TS value in the Fermi-LAT analysis is much smaller than that of SrcX, making a link between the nearby blazar and the high-energy $\gamma$-ray emission unlikely \citep{2014ApJ...788...78A,2018ApJ...861..134A,2021ApJ...911..143A}. Moreover, the distance of the blazar from VER J2016+371 and SrcX is 5.98 and 7.62 arcmin, respectively, which is much larger than their measurement uncertainties. The SrcX is separated by 1.71 arcmin away from the centroid of VER J2016+371 (near the 1$\sigma$ error radius). All these results favour the association of multiband data with CTB 87. To model the multiband data, \cite{2016MNRAS.460.3563S} also considered hadronic models to explain the observations in the GeV-TeV energy band, and found that neutral pion meson decays from proton-proton interactions with an ambient matter density of $\sim20 \ \mathrm{cm}^{-3}$ can fit the observed spectra well. However, \cite{2003ApJ...588..852K} suggested that the density may not be as high as $\sim20 \ \mathrm{cm}^{-3}$, since no limb-brightened morphology or any shell structure was found in CTB 87. Meanwhile, \cite{2018ApJ...859..173L} considered CTB 87 to be in the low-density region of the superbubble using HI 21 cm, WISE mid-IR, and optical extinction data. Therefore, the hadronic scenario appears to be an inadequate one for CTB 87, but its contribution to the total observed flux cannot be completely ruled out.

\section{summary}
\label{summary}
In this paper, an analysis of the $\gamma$-ray emission from the CTB 87 region was conducted using the latest Fermi-LAT Pass 8 data. The integral energy flux was determined to be (7.25 $\pm$ 1.36) $\times$ 10$^{-13}$ erg cm$^{-2}$ s$^{-1}$ with 5.1$\sigma$ in the 30 GeV-1 TeV energy band. The spatial position of the source is in excellent agreement with those at the radio, X-ray, and the TeV energy band. Moreover, the spectral data can be well fitted by a single power-law function with an index of 1.34 $\pm$ 0.18, which is consistent with the expectation that the PWN scenario has a hard spectrum for the GeV radiation. Combined with our GeV-TeV $\gamma$-ray data and data observed in other bands, the multiband non-thermal radiative nature of CTB 87 is investigated with the time-dependent one-zone model. With the broken power-law spectrum for the injected particles, the model can explain the observed multi-band data well with reasonable parameters. In particular, the model analysis revealed that particles injected into the nebula can be accelerated to $\sim$2.41 PeV. Further observations of CTB 87 using the LHAASO are anticipated to provide crucial insights into the origin of radiation and the mechanisms underlying particle acceleration.

\section*{Acknowledgements}
{We thank the referee for the valuable report. This work was partially supported by the National Natural Science Foundation of China (12233006, 12063004, 12393852, 12103046).
This research is supported by the Yunnan Provincial Government (YNWR-QNBJ-2018-049) and Yunnan Fundamental Research Projects (grant No. 202201BF070001-020, 202101AU070036, 202201AU070018).
}

\end{document}